\documentclass[aps,epsf,amsmath,superscriptaddress,citeautoscript,twocolumn,showpacs,floatfix]{revtex4-2}
\usepackage{bm}
\usepackage{amssymb}
\usepackage{graphicx}
\usepackage[usenames]{color}
\graphicspath{{figures/}}   %Setting the graphics path
\usepackage[normalem]{ulem} %To strike the words
\usepackage[separate-uncertainty = false,multi-part-units=single]{siunitx}
\usepackage{enumitem}
\usepackage{amsmath,amsthm,amsfonts,amscd,amssymb}
\usepackage{bm}% bold math

\begin{document}

\preprint{}

\title{Ultrafast non-equilibrium dynamics of rotons in superfluid helium}

\author{A.A. Milner}
\affiliation{Department of Physics and Astronomy, University of British Columbia, 6224 Agricultural Rd., Vancouver, B.C., Canada V6T 1Z1}

\author{P.C.E. Stamp}
\affiliation{Department of Physics and Astronomy, University of British Columbia, 6224 Agricultural Rd., Vancouver, B.C., Canada V6T 1Z1}
\affiliation{Theoretical Astrophysics, Cahill, California Institute of Technology, 1200 E. California Boulevard, MC 350-17, Pasadena CA 91125, USA}
\affiliation{Pacific Institute of Theoretical Physics, University of British Columbia, 6224 Agricultural Rd., Vancouver, B.C., Canada V6T 1Z1}

\author{V. Milner}
\affiliation{Department of Physics and Astronomy, University of British Columbia, 6224 Agricultural Rd., Vancouver, B.C., Canada V6T 1Z1}

\begin{abstract}

\end{abstract}

\maketitle

%---------------------------------------------------------------------
{\bf Superfluid $^4$He, the first superfluid ever discovered \cite{Kapitza1938, Allen1938}, is in some ways the least well understood. Unlike $^3$He superfluid \cite{ajl75, vollhardtW90}, or the variety of Bose-Einstein condensates of ultracold gases \cite{pethickS02, griffin09},  superfluid $^4$He is a very dense liquid of strongly interacting quasiparticles. The theory \cite{pinesN94, griffin02, ajl06} is then necessarily phenomenological: the quasiparticle properties are found from experiment, and controversies over their description still remain, notably regarding vortex dynamics \cite{Thouless2007, Thompson2012, Gomez2012, Sachkou2019} and the nature of rotons and roton pair creation \cite{McClintock1995}. It is therefore important to develop new experimental tools for probing the system far from equilibrium. Here we describe a method for locally perturbing the density of superfluid helium through the excitation of roton pairs with ultrashort laser pulses. By measuring the time dependence of this perturbation, we track the non-equilibrium evolution of the two-roton states on a picosecond timescale. Our results reveal an ultrafast cooling of hot roton pairs as they thermalize with the colder gas of other quasiparticles. We anticipate that these findings, as well as future applications of the introduced ultrafast laser technique to different temperature and pressure regimes in bulk liquid $^4$He, will stimulate further experimental and theoretical investigations towards better understanding of superfluidity.}

%---------------------------------------------------------------------
The dispersion of the known collective excitations in superfluid $^4$He, first proposed by Landau \cite{Landau1941a, Landau1941b}, exhibits both (i) a single quasiparticle branch with phonon character at low energy, turning over to maxons and rotons at higher energy \cite{griffin02}; and (ii) a variety of multi-quasiparticle excitations, including phonon pairs, hybridized phonon-roton excitations, roton pairs \cite{Ruvalds1970, Zawadowski1972}, and even bound roton triplets \cite{iwamoto89}. Neutron scattering \cite{godfrin18} and spontaneous Raman scattering \cite{Halley1969, Greytak1969, Greytak1970, Murray1975} give direct evidence for some of these quasiparticles. In the case of roton pairs, Raman spectra provide key information about their properties, such as the two-roton binding energy and lifetime in the thermal equilibrium state. On the other hand, the non-equilibrium quasiparticle dynamics and their approach to equilibrium are largely unknown, despite being critical for understanding the quasiparticles and their interactions. To the best of our knowledge, here we report the first experimental \textit{time-resolved} study of roton dynamics in superfluid helium.

%---------------------------------------------------------------------
In our experiment (see Methods for details), linearly polarized infrared femtosecond pulses are focused in the bulk liquid helium $^{4}$He, condensed in a custom-built optical cryostat and cooled below the transition to superfluidity (the so-called ``lambda point'' $T_{\lambda }$). The interaction of the laser field ${\bf E}({\bf r})$ with the fluid can be described by the Hamiltonian \cite{Halley1989}
\begin{equation}\label{eq-Hamiltonian}
H_\text{int}=-\int d^3r \rho({\bf r}) \cdot {\bm \mu}_\text{ind}({\bf r}) {\bf E}({\bf r}),
\end{equation}
in which ${\bm \mu}_\text{ind}({\bf r})$ is the dipole moment induced at the point ${\bf r}$ by the same field distributed over the interaction volume (points ${\bf r}'$):
\begin{equation}\label{eq-DipoleMoment}
    {\bm \mu}_\text{ind}({\bf r}) =\int d^3r' \rho ({\bf r}') \hat{\alpha} ({\bf r},{\bf r}') {\bf E}({\bf r}').
\end{equation}
Here $\rho({\bf r})$ is the density and $\hat{\alpha} ({\bf r},{\bf r}')$ is the polarizability tensor of the fluid. Owing to the dependence of the latter on the absolute value $\left|{\bf r}-{\bf r}'\right|$ \cite{Halley1989}, this interaction results in an \textit{anisotropic} re-distribution of the fluid density, and a corresponding \textit{optical birefringence}, whose optical axis is parallel to the polarization vector of the applied electromagnetic field.

We detect the laser(pump)-induced birefringence by measuring the change in polarization of a much weaker probe pulse, sent trough the liquid along the same optical path. As a function of the pump-probe delay, the birefringence signal oscillates at the frequency corresponding to the energy of a roton pair allowing us for the first time to follow the dynamics of the initially created density perturbation in the superfluid.

We note that an excited roton pair carries very little translational momentum \cite{Note1} and two units of angular momentum \cite{Udagawa1986}. Considering linearly polarized laser field as a superposition of two circularly polarized components of opposite handedness, the interaction of the field with the superfluid can also be described as a stimulated Raman process, in which the absorption of one circular component is accompanied by the emission of the other. In this regard, the observed phenomenon is analogous to the Raman-induced Kerr effect \cite{Heiman1976}, widely used in molecular spectroscopy \cite{Levenson1988}. Crucially, the stimulated mechanism of interaction provides a unique opportunity to track only those roton pairs excited by the pump pulse, rather than sampling all quasiparticles in the interaction volume (as would be the case in spontaneous Raman scattering).

%%%%%%%%%%%%%%%%%%%
\begin{figure}[t]
  \includegraphics[width=0.9\columnwidth]{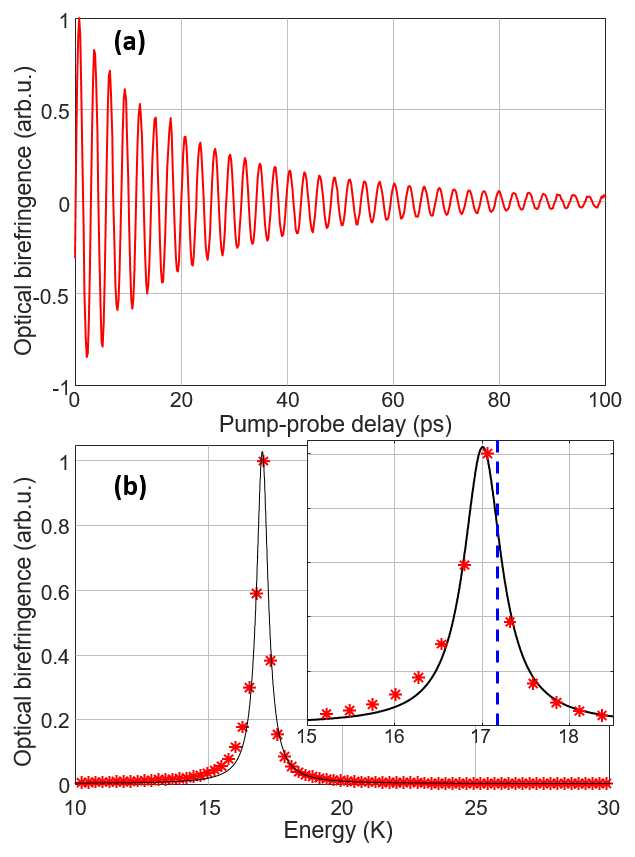}\\
  \caption{Laser-induced optical birefringence of superfluid helium at $T=\SI{1.38(2)}{K}$ and saturated vapor pressure. The raw experimental data is shown in (\textbf{a}). In (\textbf{b}), the power spectrum of the experimental signal is calculated with a Fourier transform (red asterisks) and fitted to a Lorentzian (solid black line). The details of the resonant peak are shown in the inset. The dashed vertical line marks the energy of two non-interacting rotons at this temperature and pressure.}
  \label{fig-Birefringence}
\end{figure}
%%%%%%%%%%%%%%%%%%%
%---------------------------------------------------------------------
Fig.~\ref{fig-Birefringence}(\textbf{a}) shows the oscillatory optical birefringence induced by the femtosecond laser pulse in the superfluid at a temperature $T=\SI{1.38(2)}{K}$. The oscillations decay exponentially with time, as reflected by the Lorentzian line shape of their power spectrum shown in Fig.~\ref{fig-Birefringence}(\textbf{b}).

One can see that the spectrum is dominated by a single excitation line, peaked at $\omega_{2r}(T) =\SI{17.01(1)}{K}$, which corresponds closely to the energy of a roton pair found in spontaneous Raman experiments \cite{Greytak1969, Greytak1970}. This value is slightly lower than twice the energy of a single roton, $2\Delta (T)$, measured by means of neutron scattering at the same values of temperature and saturated vapor pressure  \cite{godfrin18, Pearce2001, Fak2012}, and marked by the vertical dashed line in the inset to Fig.~\ref{fig-Birefringence}(\textbf{b}). The difference $E_b=2\Delta(T)-\omega _{2r}(T)$ is the binding energy of the two-roton state, whereas its decay rate is reflected by the measured linewidth $\gamma _{2r}(T)=\SI{0.53(1)}{K}$.

%%%%%%%%%%%%%%%%%%%
\begin{figure}[t]
  \includegraphics[width=0.95\columnwidth]{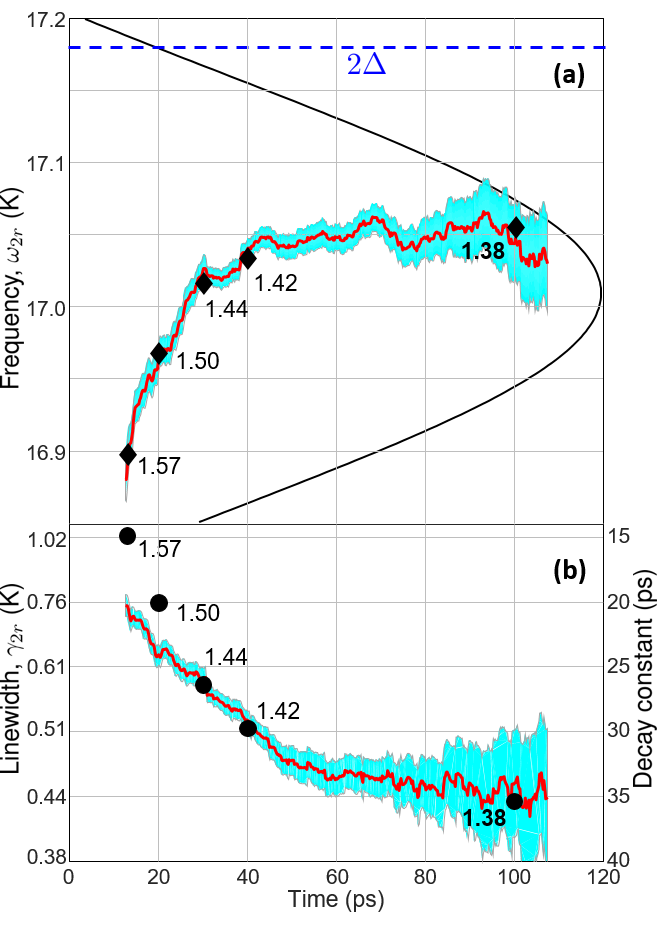}\\
  \caption{(\textbf{a}) Instantaneous frequency of the birefringence oscillations at $T=\SI{1.38(2)}{K}$ as a function of time (thick red line). See text for the calculation procedure. The dashed blue line at the top marks the energy of two free rotons $2\Delta(T=\SI{1.38}{K})$. The thin black line reproduces the top of the overall Lorentzian from Fig.~\ref{fig-Birefringence}(\textbf{b}). (\textbf{b}) Instantaneous linewidth (left scale) and the corresponding decay constant (right scale) of the birefringence oscillations at $T=\SI{1.38(2)}{K}$ as a function of time, obtained with a sliding \SI{25}{ps}-wide window. Shaded regions in both panels represent the 95\% confidence interval for the fitted frequency and linewidth, respectively. Black diamonds and circles, labeled with the corresponding temperature in K, show the comparison to the existing theory of a two-roton decay (see text for details).}
  \label{fig-EnergyShift}
\end{figure}
%%%%%%%%%%%%%%%%%%%

Examination of the spectral profile of the pump-induced birefringence shows that the line shape is skewed -- a signature of a \textit{chirped} damped harmonic oscillator \cite{Vanstaveren2010}. In comparison to a Lorentzian, the up-shifted low-energy wing of the spectrum (``red-shading'') reflects a positive frequency chirp, i.e. an increasing instantaneous frequency of the oscillator with time.

To quantify this, we ran a sliding time window across the oscillatory signal [Fig.~\ref{fig-Birefringence}(\textbf{a})], fitting the windowed part of the signal with an exponentially decaying sinusoid. The instantaneous frequency as a function of time is plotted in Fig.~\ref{fig-EnergyShift}(\textbf{a}). Here, the width of the sliding window was set at \SI{25}{ps}, which provided an optimal trade-off between the higher temporal resolution (narrower windows) and lower uncertainties of the fitting parameters (broader windows). The analysis confirms the upward frequency chirp of the observed dynamics:  the energy of the two-roton state increases during the first $\approx\SI{50}{ps}$ of its evolution, moving from the lower- to the higher-energy side of the overall line shape (thin black line).

%---------------------------------------------------------------------
Using the same sliding-window procedure, we also extract the time dependence of the two-roton linewidth $\gamma_{2r}(T)$, shown in Fig.~\ref{fig-EnergyShift}(\textbf{b}). At short times one finds $\gamma _{2r}(T)=\SI{0.75(4)}{K}$, which is significantly broader than the equilibrium value of $2 \gamma(T)=\SI{0.42}{K}$ (with $\gamma (T)$ being a single-roton decay rate at this temperature) obtained from both the neutron scattering \cite{Pearce2001} and the spontaneous Raman data \cite{Ohbayashi1998}. However, similarly to the roton pair energy, we discover that the linewidth is dynamically changing in the first \SI{50}{ps}, gradually decreasing to $\gamma_{2r}(T) = \SI{0.45(7)}{K}$, which is in good agreement with the previously known equilibrium value.

%%%%%%%%%%%%%%%%%%%
\begin{figure}[t]
  \includegraphics[width=0.95\columnwidth]{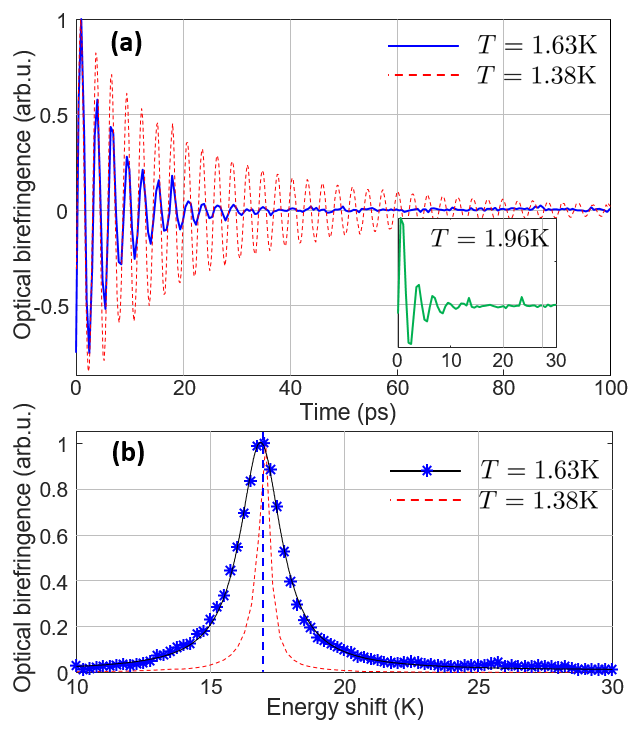}\\
  \caption{Temperature dependence of the observed birefringence signal in the time (\textbf{a}) and frequency (\textbf{b}) domains. The low-temperature data (dashed red, $T=\SI{1.38}{K}$) is the same as in Fig.~\ref{fig-Birefringence}, whereas the solid blue line correspond to $T=\SI{1.63(2)}{K}$, and the inset shows the signal at $T=\SI{1.96(2)}{K}$. The vertical blue dashed line in (\textbf{b}) marks the energy of two free rotons at that temperature, extracted from the neutron scattering data \cite{Pearce2001, Fak2012}.}
  \label{fig-TempDependence}
\end{figure}
%%%%%%%%%%%%%%%%%%%

%---------------------------------------------------------------------
As we increase the temperature of the liquid, the two-roton oscillations decay faster, as illustrated in Fig.~\ref{fig-TempDependence}(\textbf{a}), disappearing completely at the lambda point (see the inset for our highest temperature of $\SI{1.96(2)}{K}$). Worth noting are the following features:
\begin{enumerate}[label=(\roman*), leftmargin=*]
\item As seen in Fig.~\ref{fig-TempDependence}(\textbf{b}), in contrast to the lower temperature, the line shape at elevated $T$ is symmetric, and the Lorentzian profile (solid blue curve) provides a satisfactory fit to the experimental data (blue asterisks), indicating that there is no apparent frequency chirp; this is also confirmed by analyzing the power spectrum with a sliding window.
\item The line is centered at $\omega _{2r}(T)=\SI{16.87(1)}{K}$, slightly below the energy of two free rotons $2\Delta =\SI{16.97}{K}$ at this temperature (vertical dashed line).
\item Similarly to the lower temperature case, the linewidth is again considerably larger than twice the single-roton line width ($\gamma _{2r}(T)=\SI{1.88(2)}{K} \textrm{ vs } 2\gamma(T) =\SI{1.26}{K}$, with the latter found in the neutron scattering experiments \cite{Pearce2001}). Note that the spontaneous Raman data for equilibrium roton pairs at this temperature is in much better agreement with $2\gamma(T)$ (e.g. $\SI{1.19}{K}$ in \cite{Ohbayashi1998}).
\end{enumerate}

%---------------------------------------------------------------------
To understand these results, we refer to both the theoretical picture of roton line broadening, first developed by Landau and Khalatnikov \cite{Landau1949} and later extended by Bedell, Pines, Fomin and Zawadowski (BPF/BPZ) \cite{Bedell1982, Bedell1984}, and to the existing experimental data on the effect of pressure and temperature on the quasiparticle spectrum. In equilibrium, the broadening is associated with thermal roton--roton, as well as roton--phonon scattering. By fitting the experimental data from neutron \cite{Bartley1974, Gibbs1999} and spontaneous Raman \cite{Greytak1969, Greytak1970, Ohbayashi1998, Pearce2001} measurements, the phenomenological BPF/BPZ theory enables one to calculate the two-roton energy $\omega_{2r}$ and linewidth $\gamma_{2r}$ in thermal equilibrium with the superfluid at any given temperature. It shows that $\omega_{2r}$ decreases, and $\gamma_{2r}$ increases, with increasing $T$.

With this knowledge at hand, our findings immediately indicate that the roton--phonon and roton--roton interactions \textit{cool} the gas of ``hot'' laser-excited roton pairs by thermalizing them with other quasiparticles inside the interaction volume. The cooling is evidenced by the increase of $\omega_{2r}$ and the decrease of $\gamma_{2r}$ over the first \SI{50}{ps} of roton dynamics, as shown in Fig.~\ref{fig-EnergyShift}. To estimate the degree of this cooling, we chose five representative points along the measured $\omega_{2r}(t)$ curve and calculated the equilibrium temperatures, which would have resulted in the observed two-roton energies [black diamonds labeled by the corresponding temperature values in Fig.~\ref{fig-EnergyShift}(\textbf{a})]. If one assumed that the thermal equilibrium was reached at every point, the temperature of rotons must have dropped from \SI{1.57}{K} at $t=\SI{13}{ps}$ to the bath temperature of \SI{1.38}{K} at $t=\SI{100}{ps}$.

To verify whether the rapid thermalization of hot roton pairs with the colder quasiparticle gas \cite{Note2} proceeds in a quasi-equilibrium fashion, in which they can be assigned an instantaneous time-varying temperature, we employ the BPZ theory to calculate the two-roton linewidth for the (assumed) equilibrium temperatures. Comparing these calculations with our experimental results [black circles in Fig.~\ref{fig-EnergyShift}(\textbf{b})], one discovers that our quasi-equilibrium hypothesis does not necessarily hold in the first $\approx\SI{25}{ps}$, where the measured linewidths are significantly lower than the BPZ predictions.

The above interpretation can also explain the observed temperature dependence. At \SI{1.38}{K}, the cooling time is shorter than the two-roton lifetime, which enabled us to follow the thermalization of the laser-excited roton pairs prior to their decay. At higher helium temperatures, the initial roton density is higher, which increases the roton--roton collision rate and shortens the lifetime of bound roton states. As a result, at \SI{1.63}{K} roton pairs are too short-lived to equilibrate with the bath -- hence the lack of the frequency chirp in the observed $\omega _{2r}$ and the broader-than-expected linewidth $\gamma _{2r}$. The latter corresponds to the temperature of \SI{1.73}{K}, i.e. \SI{0.1}{K} hotter than the surrounding liquid, indicating that the equilibration process has not been completed. The disappearance of the two-roton signal (to within our experimental sensitivity) as we approach $T_{\lambda}$ may indicate that such excitations can exist only in the superfluid phase.

To summarize, we have shown that femtosecond laser pulses can excite density fluctuations in superfluid helium, whose dynamics are governed by laser-induced roton pairs. The fluctuations result in a measurable time-dependent optical birefringence of the fluid, which allowed us to track the evolution of the two-roton states on a picosecond timescale. This ultrafast evolution exhibits a window of non-equilibrium dynamics followed by the cooling period, during which the initially hot roton pairs equilibrate with the colder gas of phonons and rotons. Until now, access to quasiparticles in superfluid $^4$He has come from transport and thermodynamics measurements, as well as from neutron and spontaneous Raman scattering. None of these approaches can probe the short-time non-equilibrium behaviour, demonstrated in this work. Full understanding of our experimental results will require the extension of the existing theory to cover this realm.

More generally, we note that many key non-equilibrium processes in superfluids occur on ultrashort timescales (examples include the process of roton pair creation by ions in superfluid $^4$He \cite{McClintock1995}, the proposed ``baked Alaska'' scenario for B-phase nucleation in superfluid $^3$He \cite{Legett1984}, and the dynamics of vortex cores in $^4$He \cite{Thouless2007, Thompson2012, Sachkou2019}), but are hard to investigate with current methods. The experimental tool introduced here, applied to different pressures and lower temperatures, should allow us to come to grips with such open problems, and offer a much clearer picture of the non-equilibrium physics of superfluids.

%%%%%%%%%%%%%%%%%%%
\begin{figure*}[t]
  \includegraphics[width=.8\textwidth]{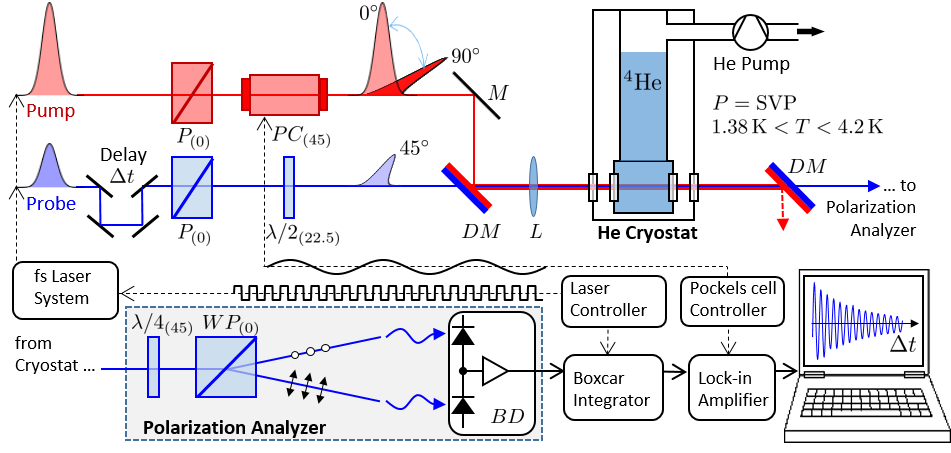}\\
  \caption{Diagram of the experimental setup. Top: femtosecond pulses with the central wavelength of 798~nm (upper, red) and 399~nm (lower, blue) serve as the pump and probe, respectively. The pulses are linearly polarized at $45^{\circ}$ with respect to one another, combined in a collinear geometry and focused into the bulk liquid helium, kept at the variable temperature and saturated vapor pressure (SVP). Bottom: after passing through the cryostat, probe pulses are filtered out from the pump light and sent to the time-gated polarization analyzer. See text for the details on the polarization modulation lock-in detection technique. $PC$: Pockels cell, $\lambda/2, \lambda/4$: zero-order half- and quarter-wave plates, $P$: polarizer, $M$: metallic mirror, $DM$: dichroic dielectric mirror, $L$:lens, $WP$: Wollaston prism (Thorlabs WP10-A), $BD$: balanced detector (Thorlabs PDB220A2). The subscripts on all polarization elements indicate the orientation of their main axis in degrees.}
  \label{fig-Setup}
\end{figure*}
%%%%%%%%%%%%%%%%%%%
%---------------------------------------------------------------------
%\begin{methods}
\subsection*{Methods}
The experimental setup is illustrated in Fig.\ref{fig-Setup}. Femtosecond pump pulses ($\approx \SI{70}{fs}$ pulse length, $\SI{1}{KHz}$ repetition rate, $\SI{798}{nm}$ central wavelength, $\SI{15}{\mu J}$ pulse energy, $\approx \SI{1e12}{W/cm^2}$ peak intensity) are focused in the bulk liquid helium, condensed in a custom-built optical cryostat (Lake Shore Cryotronics). By pumping the helium gas from the cryostat, the temperature of the liquid can be varied between $\approx\SI{1.4}{K}$ and $\SI{4.2}{K}$, while the pressure above the surface is at the saturated vapor pressure (SVP). The probe pulses ($\approx \SI{100}{fs}$ pulse length, $\SI{399}{nm}$ central wavelength) are derived from the same laser system and frequency-doubled for easier separation from the excitation light.

Our method of detecting the laser-induced linear birefringence of helium is based on an optical configuration depicted in the dashed gray rectangle in Fig.~\ref{fig-Setup}. With the pump pulses blocked, the polarization of the probe light remains unchanged all the way to the Wollaston prism ($WP$), which results in an equal split of its intensity in the two arms of the balanced detector ($BD$), and correspondingly zero signal. As soon as a small degree of linear birefringence is induced in the liquid by the pump pulses, the polarization of the probe light becomes elliptical. A quarter-wave plate [$\lambda /4_{(45)}$] converts this ellipticity into polarization rotation, shifting the balance towards one of the photo-diodes and yielding a non-zero signal.

The amplified output from the balanced detector is gated around the arrival time of the probe pulses with a Boxcar integrator. To increase the detection sensitivity, we modulate the orientation angle of the pump polarization between $0^{\circ}$ and $90^{\circ}$ at the frequency of $\SI{37}{Hz}$ by means of a Pockels cell ($PC$). Assuming that the axis of the induced birefringence follows the pump polarization, the  $BD$ signal becomes modulated at the same frequency and can be detected with a lock-in amplifier.

%\end{methods}

%\begin{addendum}
\begin{itemize}[label={}, leftmargin=*]
 \item \textbf{Acknowledgements} We are very grateful to Dr.~Apkarian and Dr.~Eloranta for many valuable discussions.
\end{itemize}
%\end{addendum}

%%%%%%%%%%%%%%%%%%%%%%%%%%%%%%%%%%%%%%%%%%%%%%%%%%%%%%%%%%%%%%%%%%%%%%%%%%%%

%\bibliography{Rotons}

\end{document}